\documentclass[11pt,a4paper]{article}
\usepackage{authblk}
\usepackage{times}
\usepackage{latexsym}
\usepackage{amsmath}
\usepackage{amssymb}
\usepackage{amsfonts}
\usepackage{graphics}
\usepackage{graphicx}
\usepackage{caption}
\usepackage{subfigure}
\usepackage{color}

\title{Van der Waals interaction between a moving nano-cylinder and a liquid thin film}

\author[1,2,3,5]{Ren\'{e} Ledesma-Alonso}
\author[2]{Elie Rapha\"{e}l}
\author[2,4]{Thomas Salez}
\author[3]{Ph. Tordjeman}
\author[3]{D. Legendre}

\affil[1]{CONACYT, Universidad de Quintana Roo, Boulevar Bah\'ia s/n, Chetumal, 77019, Quintana Roo, M\'exico.}
\affil[2]{Laboratoire de Physico-Chimie Th\'{e}orique, UMR CNRS 7083 Gulliver, ESPCI Paris, PSL Research University, 10 Rue Vauquelin, 75005 Paris, France.}
\affil[3]{Institut de M\'{e}canique des Fluides de Toulouse, UMR CNRS/INPT/UPS 5502, 2 All\'{e}e du Professeur Camille Soula, 31400 Toulouse, France.}
\affil[4]{Global Station for soft Matter, Global institution for Collaborative Research and Education, Hokkaido University, Sapporo, Hokkaido 060-0808, Japan.}
\affil[5]{\texttt{rene.ledesma@uqroo.edu.mx}}

\begin{document}
\maketitle

\abstract{
We study the static and dynamic interaction between a horizontal cylindrical nano-probe and a thin liquid film.
The effects of the physical and geometrical parameters, with a special focus on the film thickness, the probe speed, and the distance between the probe and the
free surface are analyzed.
Deformation profiles have been computed numerically from a Reynolds lubrication equation, coupled to a modified Young-Laplace equation, which takes into account the probe/liquid and the liquid/substrate non-retarded van der Waals interactions.
We have found that the film thickness and the probe speed have a significant effect on the threshold separation distance below which the jump-to-contact instability  is triggered.
These results encourage the use of horizontal cylindrical nano-probes to scan thin liquid films, in order to determine either the physical or geometrical properties of the latter, through the measurement of interaction forces.
}

\section{Introduction}

Classical Atomic Force Microscopy (AFM) experiments make possible the determination of interaction forces between nano-probes and the surfaces of liquids~\cite{OndarcuhuAime}.
In the absence of electric charges, the probe interacts with the liquid only through van der Waals (vdW) forces.
These interactions induce a deformation of the surface of the liquid~\cite{Raphael1996,Ledesma2016}.
With a probe of nanometric size, the interaction forces matter only at very small distances between the tip and the surface liquid, \textit{e.g.} a force on the order of $F=10^{-11}$ N is detected at a distance around $S=10$ nm.
However, at these distances, the force measurement becomes hard to achieve since the jump-to-contact (JTC) instability~\cite{Israelachvili2003,Ledesma2012} occurs.
In such a case, the liquid wets the probe, forming a capillary bridge~\cite{deGennes,Restagno2009} and the nature of the measured forces switches from vdW to capillary forces.
Therefore, to measure molecular forces and study the dynamics of liquid surfaces at the nanoscale, it is fundamental to estimate the critical distance, in order to approach and scan the liquid at a distance just above the JTC threshold. 

In this letter, we suggest the employment of a horizontal cylinder as a nano-probe to scan a liquid film.
The advantage of this geometry is to increase the intensity of the vdW interaction forces~\cite{Israelachvili}, while keeping a nanoscale spatial resolution in the direction perpendicular to the cylinder axis.
Comparing a nano-cylinder of radius $r$ and length $l_y$ with a spherical probe of same radius, the interaction force increases by a factor that scales as $l_y/(2r S)^{1/2}$.
For instance, a nano-cylinder, with length $l_y=1$ $\mu$m and radius $r=10$ nm at distance $S=30$ nm, generates a force $\sim40$ times stronger than a sphere of the same radius.
Alternatively, a typical interaction force of $F=10^{-11}$ N, measured by a spherical probe at $S\sim4$ nm, is also obtained with a cylindrical probe at $S\sim23$ nm, which may be a good distance to avoid the probe wetting.

Recently, we have developed a hydrodynamic model that forecasts the interaction force between a liquid and a spherical nano-probe, which also predicts the critical distance for the JTC phenomenon to occur~\cite{Ledesma2013}.
Here, we consider a nano-cylinder, which length is much larger than its radius, placed close and parallel to a liquid thin film.
We study the cylinder-liquid static interaction and the dynamic behaviour of the liquid surface due to the motion of the cylinder, perpendicular to its axis and at a constant speed.
Additionally, we show that, in both cases, the critical JTC distance decreases when the film thickness is reduced.
The threshold JTC distance is larger than that observed for a steady spherical probe.
For the cylinder-liquid dynamic case, it is found to be controlled by a critical velocity, which is a function of the film thickness.
Finally, we estimate the interaction force between the cylindrical nano-probe and the thin liquid film, for given probe speeds and separation distances. 
Using our results, one may determine some properties of the film, \textit{i.e.} its thickness or rheology, while avoiding the JTC instability.

\section{Problem formulation}

\begin{figure}
\centering
\includegraphics[width=0.48\textwidth]{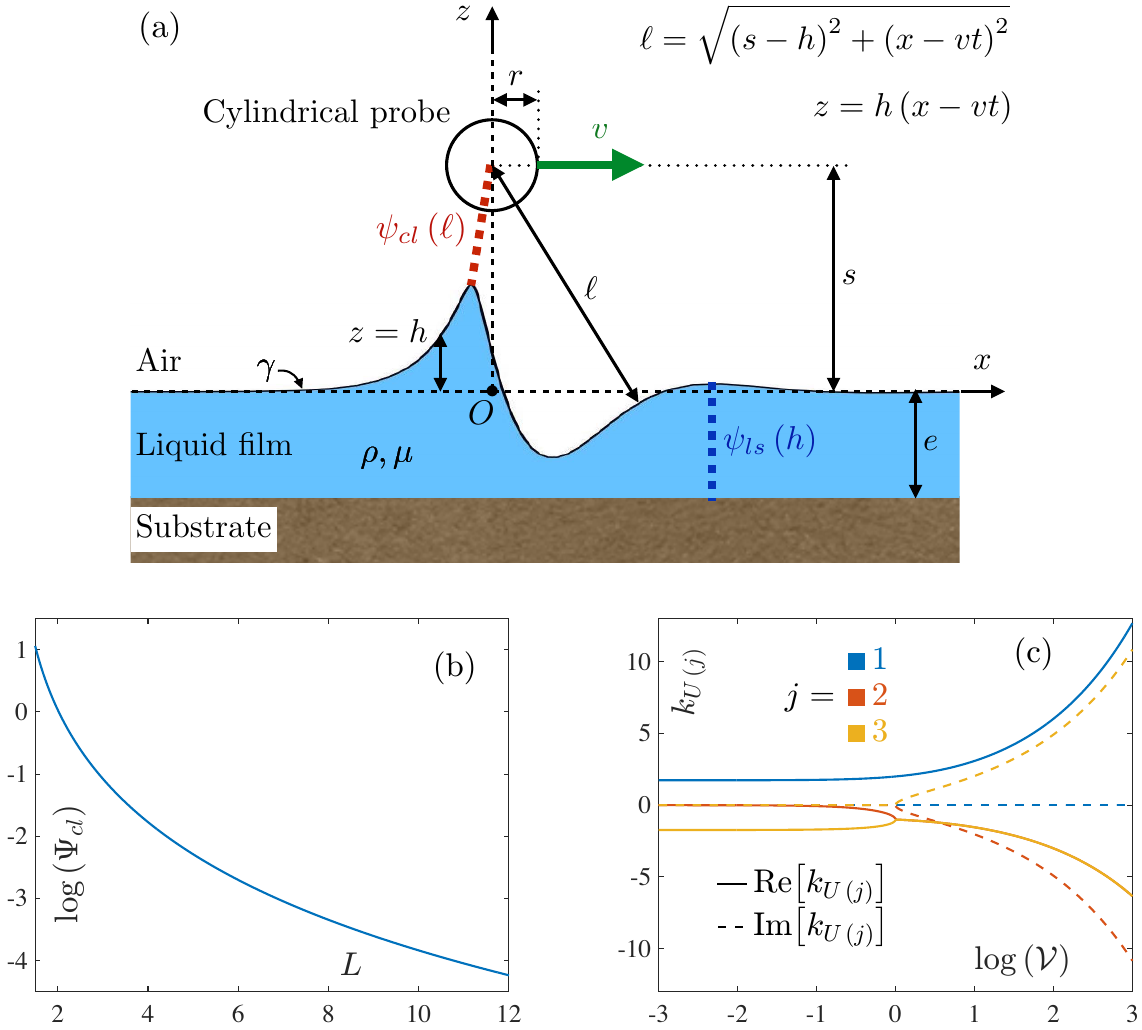}
\caption{(a) Schematic of the cylinder-liquid-substrate system, (b) dimensionless cylinder-liquid interaction potential, given by eq.~\eqref{Eq:PsiCL}, and (c) the three solutions of eq.~\eqref{Eq:Charac}, given by eq.~\eqref{Eq:CharacSol}.}
\label{Fig:Intro}
\end{figure}

We consider a liquid film of thickness $E$, density $\rho$, dynamic viscosity $\mu$ and air-liquid surface tension $\gamma$ deposited over a flat horizontal substrate.
Above the liquid film, the axis of a solid cylinder, of radius $r$ and infinitely large in the $y$-direction (see Fig.~\ref{Fig:Intro}a), is placed at a vertical distance $s$ from the flat-film surface.
The two bodies, liquid and cylinder, are attracted to each other due to the non-retarded van der Waals (nr-vdW) interaction, characterized by a Hamaker constant $A_{cl}$.
Similarly, the liquid film is also attracted to the substrate, with a Hamaker constant $A_{ls}$.
In addition, the cylinder moves parallel to the horizontal plane $z=0$, with a constant speed $v\geq 0$ in the $x$-direction.

At the air-liquid interface, the pressure difference $p$ is described by the modified Young-Laplace equation:
\begin{equation}
p=2\gamma\kappa+\rho g h-\psi_{cl}+\psi_{ls} \ ,
\label{Eq:press}
\end{equation}
where $h$ and $2\kappa$ are the position and the curvature of the liquid surface, respectively, $g$ is the acceleration of gravity, whilst $\psi_{cl}$ and $\psi_{ls}$ are the cylinder-liquid and liquid-substrate nr-vdW interaction potentials.
Moreover, lubrication theory, considering no-slip at the liquid-substrate interface and no-shear at the air-liquid interface, yields the following Reynolds equation:
\begin{equation}
\dfrac{\partial h}{\partial t}=\dfrac{\partial}{\partial x}\left[\dfrac{\left(e+h\right)^3}{3\mu}\dfrac{\partial p}{\partial x}\right] \ ,
\label{Eq:lub}
\end{equation}
which relates the dynamic behaviour of the liquid free surface to the film properties and the pressure difference.

Using the cylinder radius as a characteristic length scale, one sets the following dimensionless variables:
\begin{align}
X &=x/r \ , & \zeta &=h r/\lambda_A^2 \ , & T &=t/\tau \ , \notag \\
E &=e/r \ , & S &=s/r \ , & V &=v \tau/r \ .
\end{align}
where the Hamaker length $\lambda_A=\sqrt{A_{cl}/12\pi\gamma}$ arises from the balance between cylinder-liquid nr-vdW and surface tension forces~\cite{Stoneetal2013}.
We recall the capillary length $\lambda_C=\sqrt{\gamma/\rho g}$, yielded when surface tension and gravity effects are matched~\cite{deGennes}, the film characteristic length $\lambda_F=\sqrt{2\pi\gamma e^4/A_{ls}}$, obtained from the comparison between surface tension and liquid-substrate nr-vdW forces~\cite{Ledesma2013}, and the film viscous-capillary time $\tau=3\mu r^4/\left(\gamma e^3\right)$~\cite{deGennes}.
We also introduce the Bond number $B_o$, the Hamaker number $H_a$ and the Hamaker ratio $A$, which are defined as:
\begin{align}
B_o &=\left(r/\lambda_C\right)^2 \ , & H_a &=\left(\lambda_A/r\right)^2 \ , & A &=A_{ls}/A_{cl} \ .
\end{align}
The problem being independent of the y-coordinates, the curvature of the air-liquid interface $2\kappa$ reads, for small slopes:
\begin{align}
2\kappa &=\left(H_a/r\right) 2K \ , & 2K &=-\dfrac{\partial^2\zeta}{\partial X^2} \ .
\end{align}
where $2K$ corresponds to the dimensionless curvature.

We can also define $\ell$, the distance from the axis of the cylinder to a point (see Fig.~\ref{Fig:Intro}a) placed at the air-liquid interface with coordinates $\left(x,h\right)$, and its dimensionless equivalent $L=\ell/r$ at $\left(X,\zeta\right)$.
The distance $L$ is related to the spatial variables by the relation:
\begin{equation}
L=\sqrt{\left(S-H_a \zeta\right)^2+\left(X-V T\right)^2} \ ,
\end{equation}
Hence, the cylinder-liquid interaction potential is related to its dimensionless equivalent as follows:
\begin{equation}
\psi_{cl}=\left(\gamma H_a/r\right)\Psi_{cl} \ ,
\end{equation}
where the dimensionless potential $\Psi_{cl}$ is given by:
\begin{equation}
\Psi_{cl}=\dfrac{L+1}{\left(L^2-1\right)^3}
\left[\vphantom{\frac{1}{2}}\left(L^2+7\right) f_2-2\left(L+3\right) f_1\right] \ ,
\label{Eq:PsiCL}
\end{equation}
with $m=\left(L-1\right)/\left(L+1\right)$.
The functions $f_1\left(L\right)$ and $f_2\left(L\right)$ are defined as:
\begin{subequations}
\begin{align}
f_1 &=-\mathcal{K}\left(\sqrt{1-m^2}\right) \ , \\
f_2 &=-i \, m \, \mathcal{E}\Big(1/m\Big)+\left[1-m^2\right]f_1 \ ,
\end{align}
\end{subequations}
with $\mathcal{K}\left(z\right)$ and $\mathcal{E}\left(z\right)$ being complete elliptic integrals of the first and second kinds~\cite{EllipticK}, respectively, and $i=\sqrt{-1}$ being the imaginary unit.
The trend of the dimensionless interaction potential $\Psi_{cl}$ is shown in Fig.~\ref{Fig:Intro}b.

The liquid-substrate interaction potential is related to its dimensionless equivalent as follows:
\begin{equation}
\psi_{ls}=\left(\gamma H_a/r\right)\left(2A/E^3\right)\Psi_{ls} \ ,
\end{equation}
where the dimensionless interaction $\Psi_{ls}$ is given by:
\begin{equation}
\Psi_{ls}=1-\left(1+H_a \zeta/E\right)^{-3} \ .
\end{equation}

Finally, if we define the dimensionless pressure difference $P$ such that:
\begin{equation}
p=\left(\gamma H_a/r\right)P \ ,
\end{equation}
then eq.~\eqref{Eq:press} is re-written in dimensionless terms as:
\begin{equation}
P=2K+B_o \zeta-\Psi_{cl}+\left(2A/E^3\right)\Psi_{ls} \ ,
\label{YL:eq}
\end{equation}
and eq.~\eqref{Eq:lub} becomes:
\begin{equation}
\dfrac{\partial \zeta}{\partial T}=\dfrac{\partial}{\partial X}\left[\left(1+\dfrac{H_a \zeta}{E}\right)^3\dfrac{\partial P}{\partial X}\right] \ .
\end{equation}

\subsection{Comoving frame}
If one considers a transformation to the comoving frame of the cylinder, through the new variable $U=X-V T$, the Reynolds lubrication equation becomes:
\begin{equation}
\dfrac{\partial \zeta}{\partial T}-V\dfrac{\partial \zeta}{\partial U}=\dfrac{\partial}{\partial U}\left[\left(1+\dfrac{H_a \zeta}{E}\right)^3\dfrac{\partial P}{\partial U}\right] \ .
\end{equation}

Surface profiles that are steady in this comoving frame are obtained by setting $\partial \zeta /\partial T=0$.
Such states are identified as waves travelling to the right in the $X$-direction, with speed $V$ and without change of shape.
Therefore, for those solutions, one finds the following ODE:
\begin{equation}
\dfrac{\text{d} \zeta}{\text{d} U}=-\dfrac{1}{V}\dfrac{\text{d}}{\text{d} U}\left[\left(1+\dfrac{H_a \zeta}{E}\right)^3\dfrac{\text{d} P}{\text{d} U}\right] \ ,
\end{equation}
which, after integration and considering that $\zeta=0$ and $\text{d} P/\text{d} U=0$ at $U\rightarrow\pm\infty$, can be reduced to:
\begin{equation}
\dfrac{\text{d} P}{\text{d} U}=-V\zeta\left(1+\dfrac{H_a \zeta}{E}\right)^{-3} \ .
\label{lub:sym}
\end{equation}

\subsection{Boundary conditions}
Substituting eq.\eqref{YL:eq} within eq.\eqref{lub:sym}, considering the small-deformation $H_a \zeta/E\ll 1$ and the small-slope $H_a d \zeta/d U\ll 1$ approximations, 
and introducing the effective Bond number $B_o^{\ast}$ and the modified capillary length $\Lambda_{CF}$:
\begin{align}
B_o^{\ast} &=B_o+\left(r/\lambda_F\right)^2 \ , & \Lambda_{CF} &=r/\sqrt{B_o^{\ast}} \ ,
\end{align}
one finds the simplified equation:
\begin{equation}
  \dfrac{\text{d}^3\zeta}{\text{d} U^3}-B_o^{\ast}\dfrac{\text{d} \zeta}{\text{d} U}-V\zeta=-\dfrac{\text{d}\Psi_{cl}}{\text{d} U} \ ,
  \label{SSWE}
\end{equation}
where $\Psi_{cl}=\Psi_{cl}\left(L\right)$ and $L=L\left(U,\zeta\right)$.

For $\left\vert U\right\vert\gg S$, the dimensionless cylinder-liquid interaction and its derivative can be neglected $\text{d} \Psi_{cl}/\text{d} U\rightarrow 0$, and eq.\eqref{SSWE} reduces to:
\begin{equation}
\dfrac{\text{d}^3\zeta}{\text{d} U^3}-B_o^{\ast}\dfrac{\text{d} \zeta}{\text{d} U}-V\zeta=0 \ ,
\label{asym}
\end{equation}
which solution is given by:
\begin{equation}
\zeta=N\exp\left(\sqrt{B_o^{\ast}/3}\, k_U U\right) \ ,
\end{equation}
with $N$ being a proportionality constant.
In turn, $k_U$ is the solution of the characteristic equation:
\begin{equation}
k_U^3-3k_U-2\mathcal{V}=0 \ ,
\label{Eq:Charac}
\end{equation}
where $\mathcal{V}$ is the rescaled probe speed, defined as:
\begin{align}
\mathcal{V} &= \dfrac{v}{v_c}=\dfrac{V}{2}\left[\dfrac{3}{B_o^{\ast}}\right]^{3/2} \ , &
v_c &= \dfrac{2}{9\sqrt{3}}\dfrac{\gamma}{\mu}\left[\dfrac{e}{\Lambda_{CF}}\right]^3 \ .
\label{Eq:CharSpd}
\end{align}
Here, $v_c$ appears as a characteristic speed.
The solutions of eq.~\eqref{Eq:Charac} are:
\begin{equation}
k_{U\, (j)}=\left(\Omega/\sigma_j\right)+\left(\sigma_j/\Omega\right) \ ,
\label{Eq:CharacSol}
\end{equation}
for $j=1,2,3$, with:
\begin{align}
\Omega &=\sqrt[3]{\mathcal{V}+\sqrt{\mathcal{V}^2-1}} \ , &
\sigma_j &=\exp\left(i \dfrac{2\pi}{3} \left[j-1\right]\right) \ ,
\end{align}
where $\Omega$ satisfies the relations $\left[Re\left(\Omega\right)\right]^2=1-\left[Im\left(\Omega\right)\right]^2$ for $\mathcal{V}\leq1$ and $Im\left(\Omega\right)=0$ for $\mathcal{V}\geq1$.
The dependence of $k_{U\, (j)}$, for $j=1,2,3$, on the rescaled probe speed $\mathcal{V}$ is presented in Fig.~\ref{Fig:Intro}c.

On one hand, the asymptotic solution for $U\rightarrow -\infty$:
\begin{equation}
\zeta=N_1\exp\left(\sqrt{B_o^{\ast}/3}\, k_{U\, (1)}U\right) \ ,
\label{Sol:Left}
\end{equation}
is found, allowing us to derive the following conditions:
\begin{align}
\dfrac{\text{d}\zeta}{\text{d} U} &=\sqrt{B_o^{\ast}/3}\, k_{U\, (1)}\zeta \ , &
P &=-\sqrt{B_o^{\ast}/3}\, \dfrac{V\zeta}{k_{U\, (1)}} \ .
\label{BC:Left}
\end{align}
On the other hand, we find for $U\rightarrow \infty$:
\begin{equation}
\zeta=\sum _{j=2}^3 N_j\exp\left(\sqrt{B_o^{\ast}/3}\, k_{U\, (j)}U\right) \ ,
\label{Sol:Right}
\end{equation}
\begin{equation}
P=-\dfrac{3}{B_o^{\ast}}\dfrac{V}{k_{U\, (2)}k_{U\, (3)}}\left\{\sqrt{\dfrac{B_o^{\ast}}{3}}\left[k_{U\, (2)}+k_{U\, (3)}\right]\zeta-\dfrac{\text{d}\zeta}{\text{d} U}\right\} \ .
\label{BC:Right}
\end{equation}
Equations~(\ref{Sol:Left}-\ref{BC:Right}) are employed as boundary conditions (BCs) to find a numerical solution for the system formed by eqs.~\eqref{YL:eq} and \eqref{lub:sym}. 


\begin{table}
  \small
  \centering
  \begin{tabular*}{0.45\textwidth}{@{\extracolsep{\fill} }lc}
    Parameter & Value \\
    \hline
    Hamaker number, $H_a$ & $2.7\times10^{-3}$ \\
    Hamaker ratio, $A$ & $1$ \\
    Bond number, $B_o$ & $10^{-10}$ \\
    effective Bond number, $B_o^{\ast}$ & $\left[10^{-10} \ , \ 1.6\times10^{-2}\right]$ \\
  \end{tabular*}
  \caption{Dimensionless parameters employed in this study.}
  \label{Param:Dimless}
\end{table}

\section{Static surface profiles}

\begin{figure}
\centering
\includegraphics[width=0.49\textwidth]{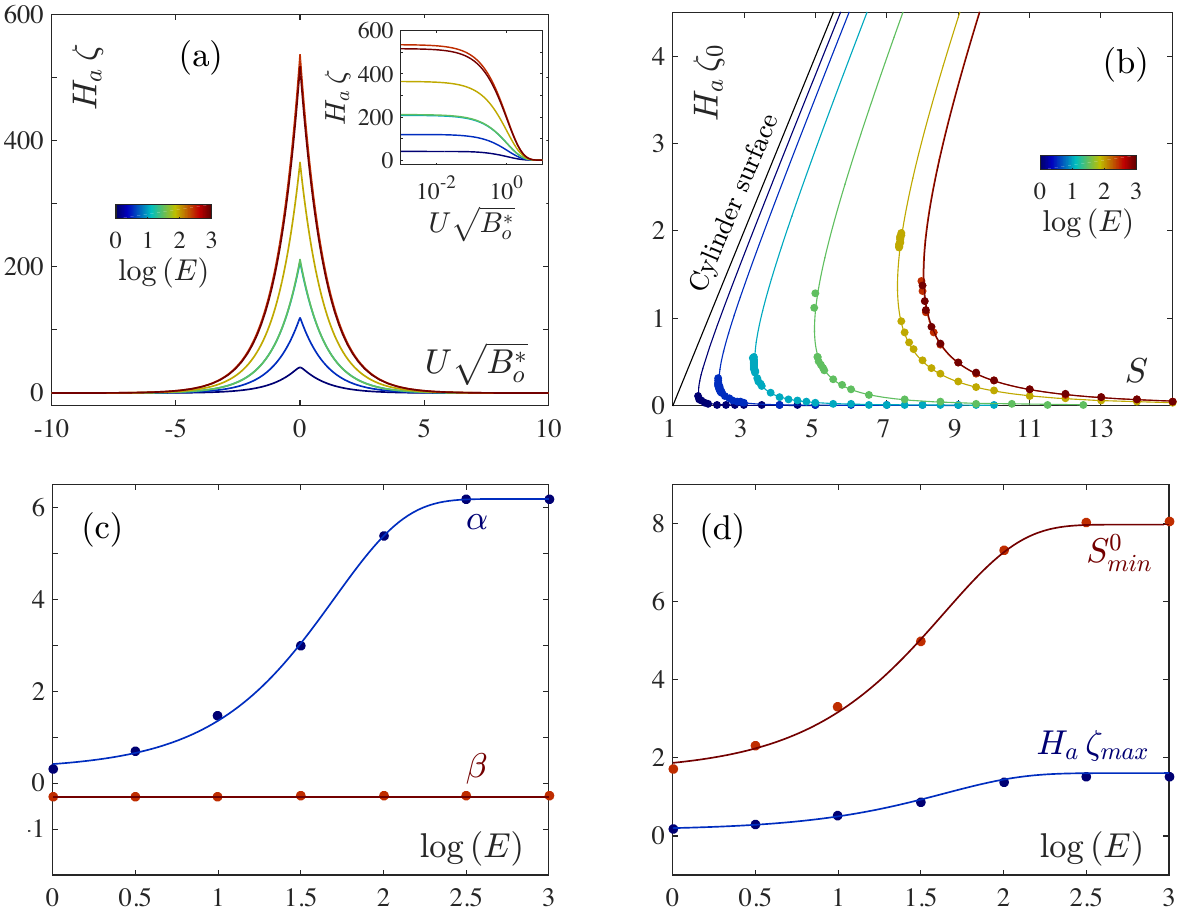}
\caption{ (a) Surface profiles for different values of the minimum separation distance $S_{min}^0$, given by the corresponding dimensionless thickness $E$ (in colors). The inset shows the $U>0$ side in $\log-\log$ scale.
(b) Apex position $\zeta_0$ as a function of the separation distance $S$, for different $E$ (in colors). Lines correspond to eq.~\eqref{Eq:Szmin}.
(c) Parameters $\alpha$ and $\beta$ as functions of $E$. The two lines are $\alpha$ given by a saturating exponential (see text) and $\beta=-0.3$.
(d) Maximum apex position $\zeta_{max}$ and $S_{min}^0$ as functions of $E$. Lines correspond to eqs.~(\ref{Eq:Szmin}--\ref{Eq:Zmax}).}
\label{Fig:Static}
\end{figure}

The static case corresponds to a situation for which the cylinder displacement speed is set to $v=0$.
This is accomplished by solving eq.~\eqref{YL:eq}, setting $P=0$, also considering the corresponding BCs and using the corresponding dimensionless parameters reported in Table~\ref{Param:Dimless}.
In Fig.~\ref{Fig:Static}a, the shape of the liquid surface $\zeta\left(U\right)$ is plotted for different values of the dimensionless film thickness, which has been varied in the range $E\in\left[10^0,10^3\right]$.
All the surface profiles present a symmetric shape, with respect to the $U=0$ axis, an exponential decay and a bump-like rounded summit with finite curvature.
The main difference resides in their amplitude, since each curve has been obtained for a particular distance $S=S_{min}^0$, whose value depends specifically on $E$.
$S_{min}^0$ corresponds to the minimum separation distance before the JTC phenomenon occurs.
In Fig.~\ref{Fig:Static}b, the apex position of the surface $\zeta_0$, which is placed at $U=0$, is presented as a function of the distance $S$.
For a given film thickness $E$, decreasing $S$ from $\infty$ towards shorter values leads to a monotonic increase of $\zeta_0$.
The probe-liquid interaction increases, pulling up the liquid surface with an increasing strength, which is consistently opposed by the surface tension, the hydrostatic and the liquid-substrate disjoining pressures, leading to an equilibrium surface profile.
There are two values of the surface apex position $\zeta_0$ at a given distance $S$, the smaller belonging to a low energy and stable branch, whereas the higher resides on a high energy and unstable branch.
At the distance $S=S_{min}^0$, the two branches connect and yield a unique equilibrium surface profile with the maximum amplitude physically possible, which corresponds to the curves shown in Fig.~\ref{Fig:Static}a for the selected values of $E$.
For shorter separation distances, below $S=S_{min}^0$, the surface tension, hydrostatic and liquid-substrate interaction effects cannot hold the strength of the probe-liquid interaction,  and the bump-like shape of the film surface becomes unstable, provoking the jump of the liquid onto the probe (JTC phenomenon) and the formation of a capillary bridge \cite{Israelachvili2003,Restagno2009}.

Due to the complexity of the cylinder-liquid interaction potential, we are not able to find analytically a simple expression to relate $\zeta_0$ and $S$.
Nevertheless, thanks to past experience~\cite{Ledesma2013}, we know that the aforementioned relation should take the form of the following Ansatz:
\begin{equation}
S\approx1+H_a\zeta_0+\alpha\left(H_a\zeta_0\right)^{\beta} \ .
\label{Eq:Szmin}
\end{equation}
where $\alpha$ and $\beta$ are parameters, whose behaviours have been determined by applying a fit to the curves shown in Fig.~\ref{Fig:Static}b.
It results that $\beta=-0.3$ is a constant, whereas $\alpha$ is a function of the film thickness $E$, which can be accurately described by the saturating exponential function $\alpha=6.2-5.9\exp\left(-0.02 E\right)$.
The parameters $\alpha$ and $\beta$ are presented in Fig.~\ref{Fig:Static}c as functions of $E$.
Also, by making $\text{d} S/\text{d} \zeta_0=0$ in eq.~\eqref{Eq:Szmin}, we find:
\begin{equation}
H_a\zeta_{max}=\left(-\alpha \beta\right)^{1/\left(1-\beta\right)} \ ,
\label{Eq:Zmax}
\end{equation}
the maximum physically possible position of the surface apex, and we are able to calculate $S_{min}^0$ by making $\zeta_0=\zeta_{max}$ in eq.~\eqref{Eq:Szmin}.
The values of $\zeta_{max}$ and $S_{min}^0$, both obtained numerically and from the combination of eqs.~(\ref{Eq:Szmin}--\ref{Eq:Zmax}), are presented in Fig.~\ref{Fig:Static}d as functions of $E$.
As the dimensionless film thickness $E$ increases, both critical values $\zeta_{max}$ and $S_{min}^0$ grow monotonically, from $\zeta_{max}=0$ and $S_{min}^0=0$ at $E=0$ towards the corresponding bulk plateau for each quantity $\zeta_{max}\left(E\rightarrow\infty\right)$ and $S_{min}^0\left(E\rightarrow\infty\right)$, which starts around $E=316$.

\section{Dynamic surface profile}

\begin{figure}
\centering
\vspace{1mm}
\includegraphics[width=0.49\textwidth]{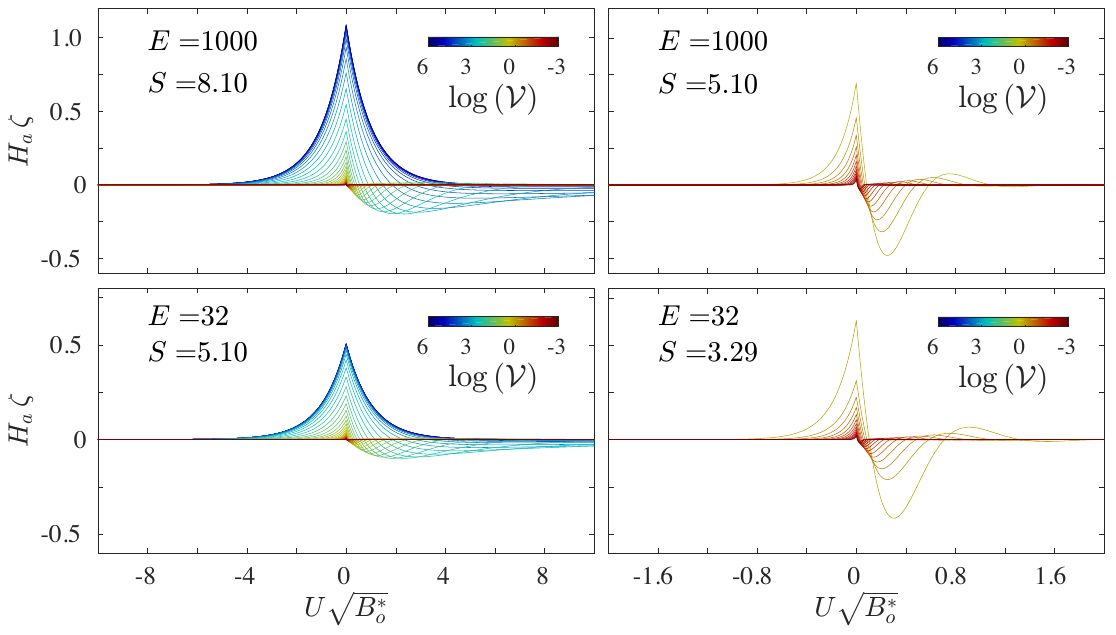}
\caption{Air-liquid interface profiles, for different values of the rescaled probe speed $\mathcal{V}$ (in colors); two film thicknesses: top row $E=1000$ (for which $S_{min}^0\approx8.04$) and bottom row $E=32$ (for which $S_{min}^0\approx4.97$); and different values of the separation distance $S$: left column $S>S_{min}^0$ and right column $S<S_{min}^0$. The probe moves from left to right.}
\label{Fig:UZVrel}
\end{figure}

In Fig.~\ref{Fig:UZVrel}, typical dynamic profiles of the air-liquid interface are shown.
They have been calculated from eqs.~\eqref{YL:eq} and \eqref{lub:sym}, with the dimensionless parameters reported in Table~\ref{Param:Dimless}, for two dimensionless film thicknesses: $E=1000$ (top row in Fig.~\ref{Fig:UZVrel}) and $E=32$ (bottom row in Fig.~\ref{Fig:UZVrel}).
For each thickness, two different values of the dimensionless distance $S$ are presented, the first (left) being $S>S_{min}^0$ and the second (right) being $S<S_{min}^0$, \textit{i.e.} above and below the static critical distance $S_{min}^0$ introduced previously.
Additionally, the rescaled probe speed has been varied in the range $\mathcal{V}\in\left[10^{-3},10^{6}\right]$.
When $S>S_{min}^0$ (left column in Fig.~\ref{Fig:UZVrel}), the effect of $\mathcal{V}$ is directly observed on the height of the apex $\zeta_{max}$.
Note that we define $\zeta_{max}$ as the highest position of the dynamic surface, which may not be placed at $U=0$ as for the static apex $\zeta_0$. 
As $\mathcal{V}$ increases, $\zeta_{max}$ decreases, together with the extent of the surface profile.
For $\mathcal{V}\leq10^{-3}$, the surface is still vertically displaced in the far field, near $\vert U\sqrt{B_o^{\ast}}\vert\sim6$, and the surface profile shows a symmetric shape with respect to the position $U=0$.
As $\mathcal{V}$ is increased, $\zeta_{max}$ lowers monotonically and the surface profile becomes asymmetric: an exponential decay for $U<0$ and oscillations within an exponential decay envelope for $U>0$.
Indeed, when the probe moves slowly, for instance $\mathcal{V}<10^{-2}$, the film has time to drain a significant amount of liquid from far-away regions towards the location of the probe, creating a nearly symmetric and high bump.
In contrast, when the probe motion is relatively fast $\mathcal{V}\in\left[10^{-2},10^0\right]$, also due to the mass conservation, the film has only time to take the liquid that is nearest to the probe, creating a sunken region in front of the probe (downstream $U>0$).
The higher the speed, the shorter the amount of collected liquid becomes, and the bump below the probe is smaller and slightly left behind to the upstream region $U<0$.
In other words, as the rescaled speed is increased above $\mathcal{V}>10^{0}$, the bump and the sunken region have less time to be formed, showing smaller magnitudes and being confined to a narrower region around $U=0$.
For $\mathcal{V}\geq10^{6}$, the surface profile is only a very small crease at $U=0$. 

When the probe speed $v$ is compared with the film characteristic speed $v_c$, which corresponds to study the relative value of $\mathcal{V}$ according to eq.~\eqref{Eq:CharSpd}, a transition from a symmetric profile towards an asymmetric behaviour is theoretically predicted.
For $\mathcal{V}<10^{-1}$ a quasi-static film profile, symmetric with an exponential decay, is found, whereas for $\mathcal{V}>10^{0}$ a non-symmetric profile, with an exponential decay at the region $U\rightarrow -\infty$ and attenuated surface oscillations at $U\rightarrow \infty$, occurs (recalling eqs.~\eqref{Sol:Left} and \eqref{Sol:Right} and the trends of $k_{U\, (j)}$, shown in Fig.~\ref{Fig:Intro}c).

For $S<S_{min}^0$  (right column in Fig.~\ref{Fig:UZVrel}), even though this range of distances corresponds to situations for which it is not possible to find a static surface profile other than a capillary bridge, non-contact dynamic profiles exist at relatively high rescaled speeds $\mathcal{V}>\mathcal{V}_{crit}$.
For a given distance $S<S_{min}^0$, the critical value $\mathcal{V}_{crit}$ corresponds to the probe speed above which we can still displace the probe along the film surface without creating a capillary bridge.
Thus, for $\mathcal{V}>\mathcal{V}_{crit}$, a non-contact profile exists (different from the capillary bridge), corresponding to an asymmetric profile, \textit{i.e.} an exponential decay for $U<0$ and oscillations within an exponential envelope for $U>0$.
A further increase of $\mathcal{V}$, provokes a reduction of the surface perturbation extent, both in the vertical and the horizontal directions, while maintaining the same shape of the decay-oscillating profile.

\begin{figure}
\centering
\includegraphics[width=0.49\textwidth]{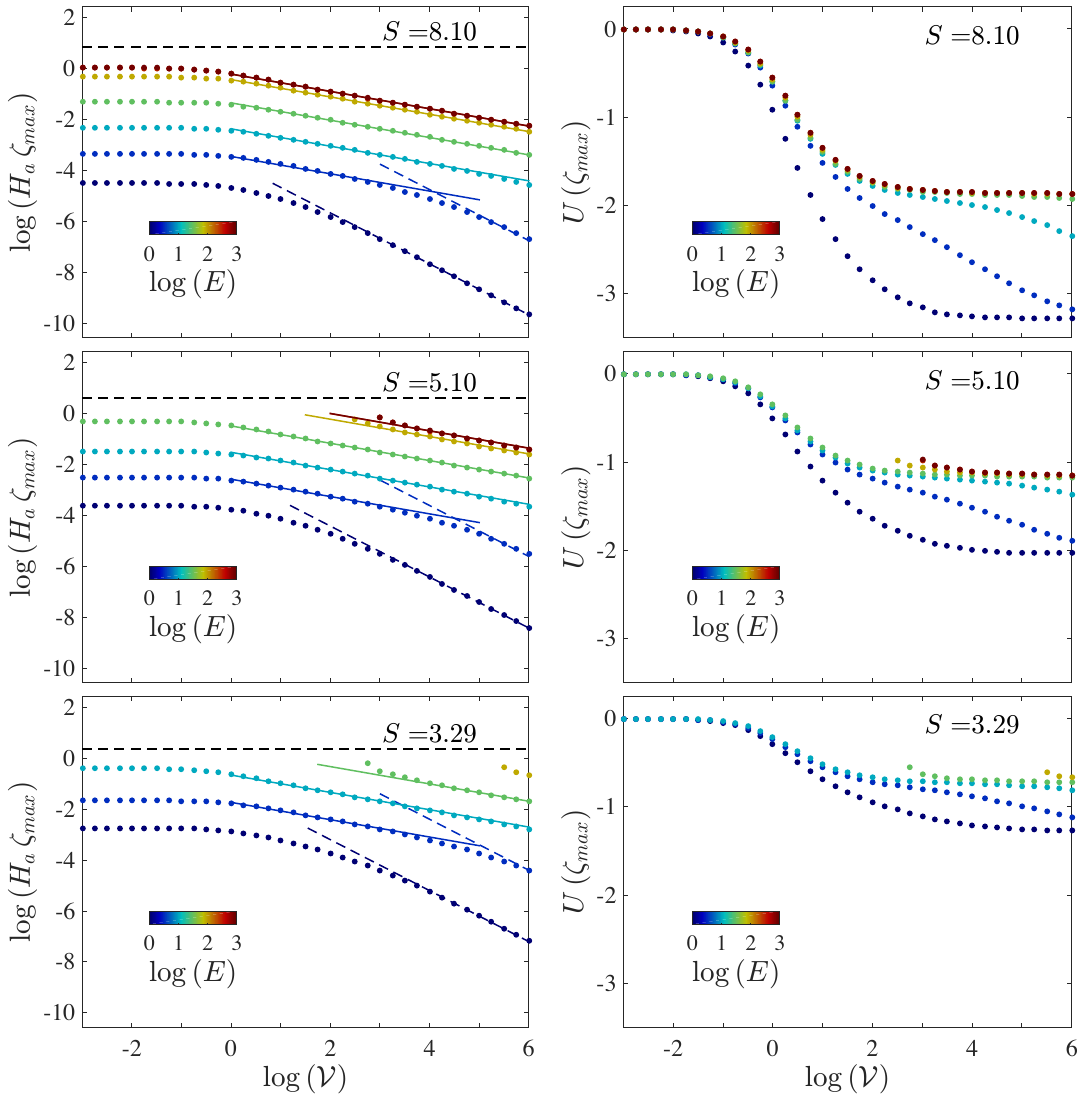}
\caption{Highest position of the film surface $\zeta_{max}$ (left column) and its lateral position $U$ (right column) as functions of the rescaled probe speed $\mathcal{V}$, for different values of the dimensionless thickness $E$ (in colors) and a given separation distance $S$. Each row corresponds to a different value of $S$: (top row) $S>S_{min}^0$ for any $E$, (middle row) $S>S_{min}^0$ for $E<100$ but $S<S_{min}^0$ for $E\geq100$, and (bottom row) $S>S_{min}^0$ for $E<32$ but $S<S_{min}^0$ for $E\geq32$. At the left column, continuous and dashed lines indicate slopes of $-1/3$ and $-1$, respectively.}
\label{Fig:ZvsV}
\end{figure}

In Fig.~\ref{Fig:ZvsV}, the surface apex $\zeta_{max}$ and its horizontal position $U\left(\zeta_{max}\right)$ are shown as functions of the rescaled probe speed $\mathcal{V}$.
They have been calculated for the dimensionless parameters reported in Table~\ref{Param:Dimless}, with the effective Bond number in the indicated range, since the film thickness has been varied within $E\in\left[10^{-3},10^6\right]$.
The separation distances $S=8.1$, $S=5.1$ and $S=3.29$, which were shown in Fig.~\ref{Fig:UZVrel}, are also the ones presented in Figs.~\ref{Fig:ZvsV}.
One should keep in mind that, for the three cases, $S$ may be larger or smaller than $S_{min}^0$, depending on the specific value of $E$ under analysis.

When $S\geq S_{min}^0$, $\zeta_{max}$ is nearly constant in the low-speed regime where $\mathcal{V}<1$, whereas its value drops when $\mathcal{V}$ is increased above $\mathcal{V}=1$.
In the high-speed regimes where $\mathcal{V}>1$, the apex position scales as $\zeta_{max}\sim\mathcal{V}^{-1/3}$ for thick films with $E>10$, whereas the scaling $\zeta_{max}\sim\mathcal{V}^{-1}$ is clearly discerned for thin films with $E=1$.
For films of intermediate thicknesses with $1<E\leq 10$, we observe the transit from a thick-film behaviour at very high speeds, to a thin-film behaviour at moderately high speeds. 
This crossover occurs at larger values of $\mathcal{V}$ as the distance $S$ is shortened.

Furthermore, still considering that $S\geq S_{min}^0$, the horizontal position $U$ of the apex $\zeta_{max}$ shifts from the probe position towards a downstream position where $U<0$, as $\mathcal{V}$ is increased.
Specifically, in the slow regime where $\mathcal{V}<1$, $U\left(\zeta_{max}\right)$ stands on the plateau $U=0$ and, at high speed where $\mathcal{V}\gg1$, it drops to either reach directly a second plateau, for thick films or thin films, or to transit slowly between them, for intermediate thicknesses.
The level of the second plateau, at high speed where $\mathcal{V}\gg1$, for both thick and thin films, gets closer to the center $U=0$ as the separation distance $S$ is diminished.

Finally, when the separation distance is shorter than the static threshold ($S< S_{min}^0$), non-contact surface profiles are only observed for a restricted speed range: $\mathcal{V}>\mathcal{V}_{crit}$.
The data points for $\zeta_{max}$ and $U\left(\zeta_{max}\right)$ are obtained in the corresponding speed regime, and are represented in Fig.~\ref{Fig:ZvsV} as left-truncated data.
As it can be observed, at $\mathcal{V}=\mathcal{V}_{crit}$, $\zeta_{max}$ diverges and $U\left(\zeta_{max}\right)$ approaches $U=0$.
When the rescaled speed $\mathcal{V}$ is increased, both quantities $\zeta_{max}$ and $U\left(\zeta_{max}\right)$ show a decreasing behaviour similar to that observed for $S\geq S_{min}^0$.

\section{Dynamic jump-to-contact distance}

\begin{figure}
\centering
\includegraphics[width=0.49\textwidth]{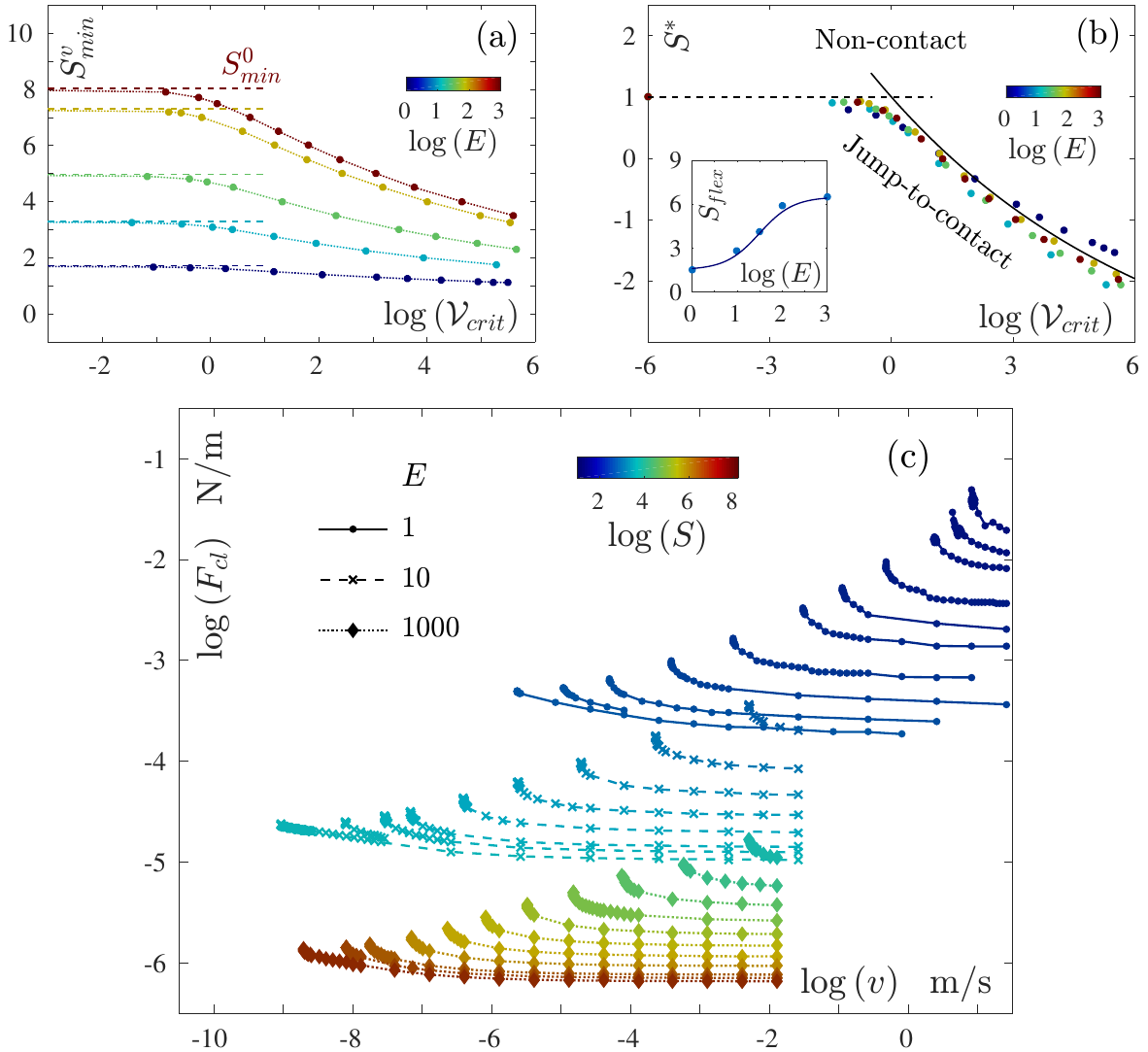}
\caption{(a) Minimum separation distance $S_{min}^v$ as a function of the critical speed $\mathcal{V}_{crit}$, for different values of the dimensionless thickness $E$ (in colors). Dashed lines represent the values of $S_{min}^0$ in the static limit $\mathcal{V}=0$, for each $E$.
(b) Phase diagram which interface is given by the reduced separation distance $S^{\ast}$, defined in eq.~\eqref{Eq:Sred}, and $\mathcal{V}_{crit}$.
The continuous line corresponds to the JTC threshold, given by eq.~\eqref{Eq:Sast}, and the dashed line represents the static value $S_{min}^0$.
The inset shows the separation distance of the inflection point $S_{flex}$, for each curve in (a), as a function of $E$. The continuous line corresponds to eq.~\eqref{Eq:Sflex}.
(c) Probe-liquid force $F_{cl}$ per unit length as a function of the probe speed $v$, for different values of $S$ and $E$.
}
\label{Fig:VrelSmin}
\end{figure}

In order to retrieve the critical speed $\mathcal{V}_{crit}$, for a film of a certain thickness $E$ and a cylinder placed at a fixed separation distance $S$, we have implemented the following procedure.
With a fixed $S$, starting from the highest value of the rescaled speed used in this study, $\mathcal{V}=10^6$, a decrease of $\mathcal{V}$ is performed.
For a distance larger than the static threshold $S>S_{min}^0$, a non-contact solution can be found for any probe speed $\mathcal{V}$.
In contrast, for $S<S_{min}^0$, a solution can be found only for speeds above the critical speed $\mathcal{V}_{crit}$.
The slope of the curve given by $\zeta_{max}\left(\log\left[\mathcal{V}\right]\right)$ is tracked, until it reaches the value of $10^3$, which we decided to be the indicator for the speed threshold, and the wetting of the probe.
This criterion also defines a minimum separation distance $S_{min}^v$ for a dynamic situation with a finite speed $\mathcal{V}>0$. 
The results of this procedure are presented in Fig.~\ref{Fig:VrelSmin}a, where the dynamic minimum distance $S^v_{min}$ is plotted against the critical speed $\mathcal{V}_{crit}$.
All the curves $S^v_{min}\left(\mathcal{V}_{crit}\right)$ follow the same trend, for low speeds $\mathcal{V}<0$ the critical distance remains at the static threshold $S^v_{min}=S^0_{min}$ and, as the speed $\mathcal{V}$ increases, $S^v_{min}$ lowers monotonically, with an inflection point occuring at a distance $S=S_{flex}$.
Using this information, a reduced separation distance $S^{\ast}$ may be defined as:
\begin{equation}
S^{\ast}=\left(S-S_{flex}\right)/\left(S_{min}^0-S_{flex}\right) \ ,
\label{Eq:Sred}
\end{equation}
which allows the data shown in Fig.~\ref{Fig:VrelSmin}a to collapse into a single curve, as presented in Fig.~\ref{Fig:VrelSmin}b.
Additionally, in Fig.~\ref{Fig:VrelSmin}b (inset), the inflection distance $S_{flex}$ as a function of $E$ is reported, for which the following empirical sigmoidal shape describes the data accurately:
\begin{equation}
S_{flex}=4+2.5\tanh\left(1.33\log\left(E\right)-2\right) \ .
\label{Eq:Sflex}
\end{equation}

Hence, one can identify the non-contact region and the dynamic jump-to-contact region, in between which the curve $S^{\ast}\left(\mathcal{V}_{crit}\right)$ acts as a boundary.
This boundary can be empirically described by the function:
\begin{equation}
S^{\ast}=5\exp\left[-0.15\log\left(\mathcal{V}_{crit}\right)\right]-4\ ,
\label{Eq:Sast}
\end{equation}
as shown in Fig.~\ref{Fig:VrelSmin}b.
For a fixed value of the reduced separation distance $S^{\ast}<1$, a rescaled speed above the threshold value $\mathcal{V}_{crit}$ allows us to scan the liquid surface without wetting the cylinder.
On the other hand, a rescaled speed below the threshold value $\mathcal{V}_{crit}$ provokes the wetting of the cylindrical probe.

\section{Force estimate}
Finally, in order to compare with experimental data, we will find the relation between force and speed, recovering their proper dimensions for direct application.
For small slopes of the air-liquid interface, the force $F_{cl}$ per unit length, mutually exerted between the cylindrical probe and the liquid film, can be approximated by~\cite{Israelachvili}:
\begin{equation}
F_{cl}=A_{cl}/\left[8\sqrt{2}r^2\left(L_{min}-1\right)\right] \ ,
\end{equation}
where $L_{min}=\sqrt{\left[S-H_a\zeta_{max}\right]^2+\left[U\left(\zeta_{max}\right)\right]^2} $ is the shortest distance between the surface of the cylinder and the free surface of the film.
Using the data obtained for $\mathcal{V}>\mathcal{V}_{crit}$, $F_{cl}$ has been computed, and is reported as a function of the probe speed $v$ in Fig.~\ref{Fig:VrelSmin}c, for the film thicknesses $e=r,10\, r,1000\, r$, each curve corresponding to a single value of the distance $s$.
For given values of $e$ and $s$, a cylindrical nano-probe scanning a thin film may experience an increasing interaction force $F_{cl}$ when the speed $v$ is lowered.
The intensity of $F_{cl}$ grows dramatically as $v$ decreases and approaches its ``no return'' value $v=v_c\mathcal{V}_{crit}$, at which the force diverges and the JTC instability is triggered. 
It is important to notice that, in Fig.~\ref{Fig:VrelSmin}c, the magnitude of the force is larger for  a thin film than for a thick film, because the threshold separation distances is shorter for the former than for the latter.
It is, in consequence, the combination of thickness $e$ and distance $s$ which defines the force magnitude for a given speed $v$, as i can be discerned in Fig.~\ref{Fig:VrelSmin}c.

\section{Conclusions}

We have studied the effects of the non-retarded van der Waals interaction between a moving cylindrical probe and a viscous thin film deposited over a rigid substrate.
The influences of the physical and geometric parameters have been analyzed via a few dimensionless parameters: effective Bond number $B_o^{\ast}$, Hamaker number $H_a$, separation distance $S$ and film thickness $E$.
We have found that for both static and dynamic situations, the amplitude of the deformation increases with the Hamaker number $H_a$, but decreases with the effective Bond number $B_o^{\ast}$.
In addition, we have verified that shortening the separation distance $S$ leads to larger displacements $\zeta_{max}$ of the free-surface profile.

Another dimensionless parameter, the rescaled probe speed $\mathcal{V}$, plays a major role in the dynamic phenomenon.
This new parameter controls the morphology of the film surface.
Low speeds $\mathcal{V}<10^{-1}$ yield a quasi-static surface profile, symmetric with respect to the horizontal position of the cylindrical probe $U=0$, with an exponential-decay length proportional to $\left(B_o^{\ast}\right)^{-1/2}$.
High speeds $\mathcal{V}>10^{1}$ yield an asymmetric surface profile, with an exponential-decay length proportional to $\left(B_o^{\ast}\right)^{-1/2}\mathcal{V}^{-1/3}$ at horizontal positions where $U\ll0$ and attenuated oscillations at $U\gg0$.

We have also unveiled that increasing the rescaled speed $\mathcal{V}$, above a critical value $\mathcal{V}_{crit}$, which depends on the separation distance $S$, can prevent the jump-to-contact (JTC) instability.
Alternatively, when scanning at a finite rescaled speed $\mathcal{V}>0$, the probe can be placed closer to the free surface of the liquid film, since the dynamic minimum separation distance $S^v_{min}$, below which the JTC instability is triggered, is smaller than the static threshold value $S^0_{min}$.
A phase diagram has been presented, in terms of the rescaled speed $\mathcal{V}$ and the reduced separation distance $S^{\ast}\left(S,S^0_{min},E\right)$, which allows us to identify the dynamic non-contact and jump-to-contact regions.
This result may be useful for determining the conditions to perform local-probe scanning experiments, since experimentalists may be able to increase the sensitivity of the probe by reducing the separation distance, while avoiding the wetting instability by increasing the probe speed.
Additionally, the presented methodology can be employed to determine one of the physical or geometric parameters, \emph{e.g.} film thickness or viscosity (rheology), when the remaining parameters are known.
In comparison with a spherical probe, a cylindrical probe may yield a finer estimate of the film properties, since the measurements can be performed at shorter probe/liquid distances.

\bibliographystyle{unsrt}
\bibliography{ProbeDisplBiblio}

\begin{thebibliography}{10}

\bibitem{OndarcuhuAime}
T.~Ondarcuhu and J.P. Aime.
\newblock {\em Nanoscale liquid interfaces: Wetting, patterning and force
  microscopy at the molecular scale}.
\newblock Pan Stanford Publishing, 2013.

\bibitem{Raphael1996}
E.~Raphael and P.-G. de~Gennes.
\newblock Capillary gravity waves caused by a moving disturbance: Wave
  resistance.
\newblock {\em Phys. Rev. E}, 53:3448--3455, 1996.

\bibitem{Ledesma2016}
R.~Ledesma-Alonso, M.~Benzaquen, T.~Salez, and E.~Raphael.
\newblock Wake and wave resistance on viscous thin films.
\newblock {\em J. Fluid Mech.}, 792:829--849, 2016.

\bibitem{Israelachvili2003}
N.~Maeda, J.~N. Israelachvili, and M.M. Kohonen.
\newblock Evaporation and instabilities of microscopic capillary bridges.
\newblock {\em P. Natl. Acad. Sci.}, 100(3):803--808, 2003.

\bibitem{Ledesma2012}
R.~Ledesma-Alonso, D.~Legendre, and Ph. Tordjeman.
\newblock Nanoscale deformation of a liquid surface.
\newblock {\em Phys. Rev. Lett.}, 108:106104, 2012.

\bibitem{deGennes}
P.-G. de~Gennes, F.~Brochard-Wyart, and D.~Quere.
\newblock {\em Capillarity and wetting phenomena: Drops, Bubbles, Pearls,
  Waves}.
\newblock Springer, 2003.

\bibitem{Restagno2009}
L.~Vagharchakian, F.~Restagno, and L.~Leger.
\newblock Capillary bridge formation and breakage: A test to characterize
  antiadhesive surfaces.
\newblock {\em J. Phys. Chem. B}, 113(12):3769--3775, 2009.

\bibitem{Israelachvili}
J.~N. Israelachvili.
\newblock {\em Intermolecular and surface forces}.
\newblock Elsevier, 3rd edition, 2011.

\bibitem{Ledesma2013}
R.~Ledesma-Alonso, D.~Legendre, and Ph. Tordjeman.
\newblock Afm tip effect on a thin liquid film.
\newblock {\em Langmuir}, 29:7749--7757, 2013.

\bibitem{Stoneetal2013}
D.B. Quinn, J.~Feng, and H.A. Stone.
\newblock Analytical model for the deformation of a fluid-fluid interface
  beneath an afm probe.
\newblock {\em Langmuir}, 29:1427--1434, 2013.

\bibitem{EllipticK}
E.W. Weisstein.
\newblock Elliptic integrals of the first kind \& second kind.
\newblock \footnotesize From MathWorld -- A Wolfram Web Resource.
  http://mathworld.wolfram.com.

\end{thebibliography}

\end{document}